\documentclass[11pt]{article}

\usepackage{graphicx,epsf,subfig}
\usepackage{amsmath,amsfonts, natbib,verbatim,booktabs,longtable,multirow,lscape}
\usepackage{enumerate, color, url}
\usepackage{dsfont,ctable}
\usepackage[margin = 1in]{geometry}
\usepackage{setspace}
\usepackage{algorithm}
\usepackage{wasysym}
\usepackage{makecell}

\newtheorem{theorem}{Theorem}[section]

\newtheorem{lemma}[theorem]{Lemma}

\newtheorem{assumption}[theorem]{Assumption}

\newcommand{\beq}{\begin{equation}}  % expects a label
\newcommand{\eeq}{\end{equation}}

\newcommand{\ben}{\begin{enumerate}}  % expects a label
\newcommand{\een}{\end{enumerate}}
\newcommand{\bed}{\begin{itemize}}  % expects a label
\newcommand{\eed}{\end{itemize}}

\newcommand {\bSigma} {{\boldsymbol\Sigma}}

\newcommand {\GG} {{\mathcal G}}

\newcommand{\one}{\mathds{1}}
\newcommand{\sumi}{\sum_{i=1}^n}

\newcommand {\bX} {{\mathbf{X}}}

\title{The cumulative Kolmogorov filter for model-free screening in ultrahigh dimensional data}
\author{Arlene K. H. Kim and Seung Jun Shin\\ \bigskip
\normalsize{ \it University of Cambridge and Korea University}}
\date{}

\begin{document}

\maketitle

\begin{abstract}
We propose a cumulative Kolmogorov filter to improve the fused Kolmogorov filter proposed by \cite{mai2015fused} via cumulative slicing. We establish an improved asymptotic result under relaxed assumptions and numerically demonstrate its enhanced finite sample performance.
\\ \vspace{10mm}
\end{abstract}

\section{Introduction}
Since \citet{fan2008sure}, a marginal feature screening has been regarded as one canonical tool in ultrahigh-dimensional data analysis. Let $Y$ be a univariate response and $\bX = (X_1, \ldots, X_p)^T$ be a $p$-dimensional covariate. We assume 
that only a small subset of covariates are informative to explain $Y$. In particular, we assume $|S^*| = d \ll p$ where
\begin{align}\label{eq::set}
S^* = \{j: F(y|\bX) \mbox{ functionally depends on } X_j \mbox{ for some }  y \},
\end{align}
with $F(\cdot|\bX)$ being the conditional distribution function of $Y|\bX$.
Such assumption is reasonable since including large number of variables with weak signals often deteriorates the model performance due to accumulated estimation errors.

Since the introduction of 
\citet{fan2008sure}, numerous marginal screening methods have been developed (see Section 1 of \citet{mai2015fused} for a comprehensive summary). Among these methods, model-free screening \citep{zhu2011model,li2012feature,mai2015fused} is desirable since the screening is a pre-processing procedure followed by a main statistical analysis. 

For feature selection in binary classification, Kolmogorov filter (KF) is proposed by \citet{mai2012kolmogorov}. For each $X_j, j = 1, \ldots, p$, KF computes
$$
\kappa_j = \sup_x| P(X_j \le x|Y = 1) - P(X_j \le x|Y = -1)|, \quad \mbox{$ j = 1,\ldots,p$,}
$$ 
and selects variables with large $\kappa_j$'s among all $j = 1, \cdots, p$. A sample version of $\kappa_j$ is obtained by replacing the probability measure with its empirical counterpart, leading to the well-known Kolmogorov--Smirnov statistic where its name came from. KF shows impressive performance in binary classification.

Recently, \citet{mai2015fused} have extended the idea of KF beyond the binary response by slicing data into $G$ pieces depending on the value of $Y$. In particular, a pseudo response $\tilde Y$ taking $g$ if $Y \in (a_g-1, a_g]$ for $g=1, \ldots, G$, is defined for given knots $\mathcal{G} = \{(-\infty =) a_0 < a_1 < \ldots < a_G (= \infty)\}$. Following the spirit of KF, one can select a set of variables with large values of 
\begin{equation} \label{eq:fkf}
\kappa_j^{\GG} = \max_{l,m} \sup_x | P(X_j \le x|\tilde Y = m) - P(X_j \le x|\tilde Y = l)|, \quad \mbox{$j = 1,\ldots,p$}.
\end{equation}
However, information loss is inevitable due to the lower resolution of pseudo variable $\tilde Y$ compared to $Y$ regardless of the choice of $\mathcal{G}$. To tackle this, \citet{mai2015fused} proposed fused Kolmogorov filter (FKF) that combinies $\kappa_j^\mathcal{G}$ for different $N$ sets of knots $\mathcal{G}_1, \ldots, \mathcal{G}_N$ and selects variables with large values of $\kappa_j^{\text{fused}} = \sum_{\ell = 1}^N \kappa_j^{\mathcal{G}_\ell}$, for $j = 1,\ldots, p$. The source of improvement in FKF is clear, however, it cannot perfectly overcome the information-loss problem caused by slicing. In addition, it is subtle to decide how to slice data in a finite sample case. To this end, we propose the cumulative Kolmogorov filter (CKF). CKF minimizes information loss from the slicing step and is free from choice of slices. As a consequence, it enhances the FKF.

\section{Cumulative Kolmogorov filter}
We let $F(\cdot | X_j)$ denote the conditional distribution function of $Y$ given $X_j$. Given $x$  such that $0< P(X_j \leq x)<1$, define 
\begin{equation} \label{eq::ckf1}
k_j(x) = \sup_{y} \left|F(y| X_j> x) - F(y|X_j \leq x) \right|, \qquad j = 1, \ldots, p.
\end{equation}
We remark that \eqref{eq::ckf1} is identical to \eqref{eq:fkf} with $\mathcal{G} = \{-\infty , x , \infty\}$ except that the sliced variable in \eqref{eq::ckf1} is $X_j$ instead of $Y$. The choice of a slicing variable between $X_j$ and $Y$ is not crucial, however, it would be more natural to slice independent variable in regression set up whose target is $E(Y|\bX)$. 
Now, 
$$
k_j(x) = \frac{1}{P(X_j \leq x) (1-P(X_j \leq x)) } \sup_y \left|  P(X_j \leq x) P(Y \leq y) - P(Y\leq y, X_j \leq x) \right|,
$$
which immediately yields $k_j(x) = 0$ for all $x$ satisfying $0<P(X_j \leq x) <1$ if and only if $X_j$ and $Y$ are independent. In fact, $k_j(x)$ indicates the level of dependence as shown in the following lemma.

\begin{lemma}\label{bivariate}
 If $(X_j,Y)$ has a bivariate Gaussian copula distribution such that $(g_1(X_j),g_2(Y))$ is jointly normal with correlations $\rho_j = \text{Cor}(g_1(X_j), g_2(Y))$ after transformation via two monotone funcitons $g_1,g_2$, and $g_1(X_j)$ and $g_2(Y)$ are marginally standard normal. Then 
 \begin{enumerate}
 \item $k_j(x) = 1$ if $|\rho_j| = 1$ and $k_j(x) = 0$ if $\rho_j = 0$.
  \item Denoting $y^* = x \big(\frac{1-\sqrt{1-\rho_j^2}}{\rho_j} \big)$,
  \begin{align*}
k_j(x) &= \frac{1}{\Phi(x)(1-\Phi(x))} \left|\int_{-\infty}^{y^*} \Phi \Big( \frac{x- \rho_j u}{\sqrt{1-\rho_j^2}} \Big) \phi(u) du - \Phi(x) \Phi(y^*) \right|. 
\end{align*}
\item For each $x$, $k_j(x)$ is a strictly increasing function of $|\rho_j|$.
 \end{enumerate}
\end{lemma}

Nonetheless, \eqref{eq::ckf1} loses lots of information from the dichotomization of $X_j$. To overcome this, we define
\begin{align} \label{cum.kf}
K_j = E\left[k_j(\tilde X_j)\right],\quad \mbox{for \ $j = 1, \ldots, p$,}
\end{align}
where $\tilde X_j$ denotes an independent copy of $X_j$. 
In the population level, \eqref{cum.kf} is fusing infinitely many KFs with all possible dichotomized $X_j$'s. By doing this, we can not only minimize efficiency loss but also be free from the choice of knot sets. Similar idea has been firstly proposed by \citet{zhu2010dimension} in the context of sufficient dimension reduction where the slicing scheme has been regarded as a canonical approach.

Given $(Y_i, \bX_i), i = 1, \ldots, n$ where $\bX_i = (X_{i1}, \ldots, X_{ip})^T$, a sample version of \eqref{eq::ckf1} is 
$\hat k_j(x) =  \sup_y \left| \hat F(y| X_j >x) - \hat F(y| X_j \leq x) \right|$ where $\hat F(y| X_j > x) = \frac{\sum_{i=1}^n \one_{\{Y_i \leq y, X_{ij} > x \}}}{\sum_{i=1}^n \one_{\{ X_{ij} > x\}}}$ and $\hat F(y| X_j \le x)$ is similarly defined. 
Following the convention, we regard $0/0 = 0$.
Now, an estimator of \eqref{cum.kf} is given by
\begin{align} \label{est.cum.kf}
\hat K_{j} = \frac{1}{n}\sumi \hat k_{j}(X_{ij}).
\end{align}
Finally, for $d_n \in \mathbb{N}$, we propose CKF to select the following set
$$
\hat S(d_n) = \{j: \hat K_{j} \mbox{ is among the first $d_n$ largest of all $\hat K_j, j = 1, \cdots, p$} \}.
$$

\section{The Sure Screening Property}
We assume a regularity condition.
\begin{assumption}\label{ass}
 There exists a nondegenerate set $S$ such that $S^* \subseteq S$ and 
 $$\Delta_S = \min_{j \in S} K_j - \max_{j \notin S} K_j >0.$$
\end{assumption}
Assumption \ref{ass} is similar to the regularity condition (C1) for KFK \citep{mai2015fused}. In fact, FKF requires one additional condition that guarantees that the estimated slices are not very different from oracle slices based on population quantiles of $Y$, which is not necessary for CKF since it is free from the slice choice. KF with a binary response requires only one assumption similar to Assumption~\ref{ass}.

\begin{theorem}\label{mainthm}
 Under Assumption \ref{ass}, when $d_n \geq |S|$ and $\Delta_S>4/n$, 
 $$
 P( S^* \subset \hat S(d_n)) \geq 1- \eta,
 $$
 where 
 $$
 \eta = p \left(4n\exp(-n\Delta_S^2/128) + 2\exp(-n\Delta_S^2/16) \right).
 $$
 This probability tends to 1 when $\Delta_S \gg \sqrt{\frac{\log (pn)}{n}}$.
\end{theorem}

The sure screening probability converges to one when  $\Delta_S \gg \{\log (p n)/n\}^{1/2}$.

\section{A simulation study}
\subsection{A toy example}
\begin{figure}[h]
\centering
\includegraphics[height = 6cm]{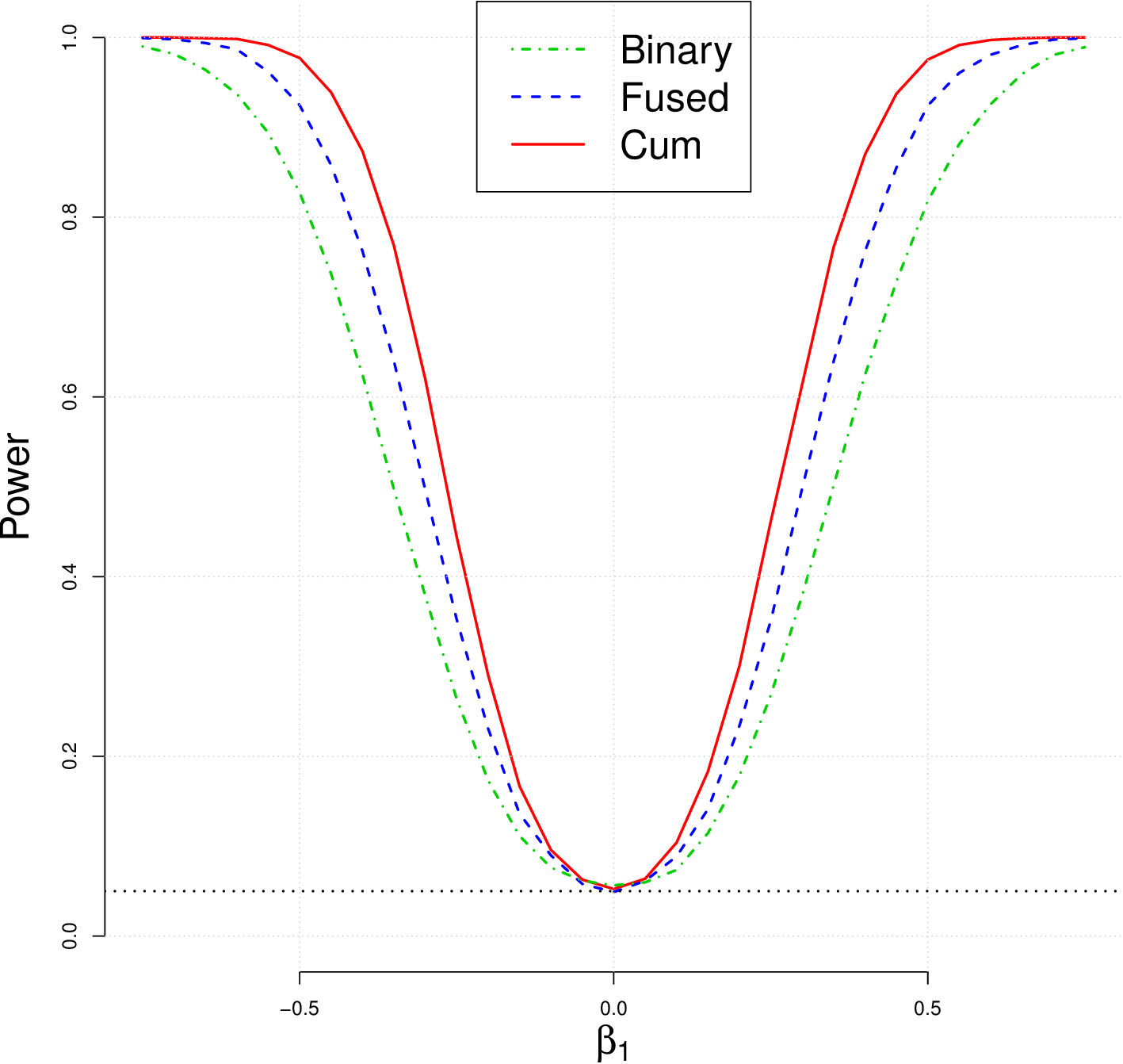}
\caption{Power functions under $\alpha = 0.05$. CKF shows clear improvement.}
\label{fg:motivation}
\end{figure}

Consider a simple regression model $Y = \beta X + \epsilon$ where $X$ and $\epsilon$ are from independent $N(0,1)$. In this regard, \eqref{est.cum.kf} can be thought as a statistic for testing $H_0: \beta = 0$. To demonstrate the performance of CKF, we compare its power to i) $\hat \kappa^{\text{binary}} = \sup_y | \hat F(y|X> \mbox{median}\{X\}) - \hat F(y|X \le \mbox{median}\{X\})|$ and ii) $\sum_{\ell = 1}^4 \hat \kappa^{\mathcal{G}_\ell}/4$ with four equally-spaced knot sets whose sizes are 3,4,5, and 6 as suggested by \citet{mai2015fused}. Figure \ref{fg:motivation} depicts numerically computed power functions of three methods under significance level $\alpha = 0.05$. As expected, CKF \eqref{est.cum.kf} performs best while the simplest $\hat \kappa^{\text{binary}}$ does worst, which echoes the fact that screening performance can be improved by minimizing information loss entailed in the slicing step and CKF indeed achieves it. 

\subsection{Comparison to other screening methods}
We consider the following nine models with $(n, p) = (200,  5000)$ and $\epsilon \sim N(0,1)$ independent of $\bX$: 
\begin{enumerate}
\item  $U(Y) = T(\bX)^T \beta + \epsilon$, where $\beta = (2.8 \times 1_2^T, 0_{p-2}^T)^T$, $T(X) \sim N_p (0_p, \bSigma)$ with $\bSigma = CS(0.7)$. $CS(0.7)$ is a compound symmetry correlation matrix with the correlation coefficient of $0.7$. Let $U(Y) = Y$, $T(\bX) = \bX$. 
\item  $T(\bX) = \bX^{1/9}$ and other settings are the same as Model 1. 
\item $U(Y) = Y^{1/9}$ and other settings are the same as Model 1. 
\item $U(Y) = T(\bX)^T \beta + \epsilon$, where $\beta = (0.8 \times 1_{10}^T, 0_{p-10}^T)^T$, $T(X) \sim N_p (0_p, \bSigma)$ with $\bSigma = AR(0.7)$. $AR(0.7)$ is an autoregressive correlation matrix with the autoregressive correlation coefficient of $0.7$. Let $U(Y) = Y$, $U(\bX) = \bX$. 
\item $T(\bX) = \frac{1}{2}\log(\bX)$ and  and other settings are the same as Model 4.
\item $U(Y) = \log(Y)$ and other settings are the same as Model 4.

\item $Y = (X_1 + X_2 + 1)^3 + \epsilon$, where $X_j \stackrel{iid}{\sim} Cauchy$. %The intrinsic dimension $d$ is $2$.
\item $Y = 4X_1 + 2 \tan (\pi X_2/2) + 5X_3 + \epsilon$,  where $X_j \stackrel{iid}{\sim} U(0,1)$ independently.
\item $Y = 2(X_1 + 0.8 X_2 + 0.6 X_3 + 0.4 X_4 + 0.2 X_5) + \exp(X_{20} + X_{21} + X_{22})\epsilon$,
 where $\bX \sim N(0, \bSigma)$ with $\bSigma = CS(0.8)$. 
\end{enumerate}

To avoid a cutoff selection problem, we report the average number of minimum variables needed to recover all  informative ones over 100 independent repetitions. Hence, a smaller value implies a better performance. Table \ref{sim1} contains the comparison results against correlation learning \citep[CS,][]{fan2008sure} and distance correlation learning \citep[DCS,][]{li2012feature} as well as FKF. The results clearly show that the proposed CKF has improved performance compared to others including FKF. 

\begin{table} \scriptsize
\centering
\begin{tabular}{c c rrrrrrrr} \hline
Model & $d$ & 
\multicolumn{2}{c}{SIS}   & \multicolumn{2}{c}{DCS} & 
\multicolumn{2}{c}{FKF} & \multicolumn{2}{c}{CKF} \\ \hline
1 &  2 &    2.00 &    (0.00) &    2.00 &    (0.00) &  3.79 &   (6.28) &  2.00 &  (0.00)\\
2 &  2 & 2038.12 & (1348.05) & 1985.10 & (1460.82) &  4.62 &   (9.14) &  2.00 &  (0.00)\\
3 &  2 &  891.22 & (1071.58) &  350.88 &  (794.67) &  3.88 &   (6.96) &  2.00 &  (0.00)\\ \hline
4 & 10 &   10.04 &    (0.20) &   10.04 &    (0.20) & 10.26 &   (1.09) & 10.06 &  (0.24)\\
5 & 10 &  150.10 &  (351.46) &   12.50 &   (10.42) & 10.23 &   (0.49) & 10.11 &  (0.35)\\
6 & 10 & 1618.50 & (1423.11) &  927.16 &  (916.20) & 10.81 &   (4.27) & 10.03 &  (0.17)\\ \hline
7 &  2 & 1051.14 & (1473.43) &  682.47 &  (965.43) &  2.00 &   (0.00) &  2.00 &  (0.00)\\
8 &  3 & 2980.23 & (1494.26) &  277.43 &  (606.47) &  9.05 &  (18.69) &  6.66 & (11.27)\\
9 &  8 & 3562.30 & (1252.76) &  231.63 &  (526.51) & 60.84 & (126.12) & 38.59 & (52.58)\\ \hline
% \hline
\end{tabular}
\caption{Average number of minimum variables needed to keep all  informative ones over 100 independent repetitions. Standard deviations are in parentheses.}\label{sim1}
\end{table}

\section{Discussions}
We employ a cumulative slicing technique to extend a screening tool for binary response to contiuous one. The idea is quite general and can be applied to t-test-based screening \citep{fan2008high,fan2008sure} as well as logistic-regression-based screening \citep{fan2010sure}. 
In addition, it is possible to extend the idea of CKF to the censored response by replacing the empirical distribution function with the Kaplan-Meier estimator.  

\bibliographystyle{dcu}%elsarticle-num}
\bibliography{references}      %%%%%

\appendix
\section{Proof of Lemma \ref{bivariate}}
Because $k_j$ is invariant under monotone transformation, it suffices to consider the case where $g_1(t) = t$, $g_2(t)=t$, and thus $X_j$ and $Y$ are jointly normal. 
%If $\rho=1$, then the supremum is attained by taking $y = x$ and if $\rho=-1$, then the supremum is attaied by taking $y=-x$ which gives $k_j(x) = 1$. 
If $\rho_j=0$, then $X_j$ is independent of $Y$ and $k_j(x) = 0$. On the other hand, if $\rho_j \neq 0$, $Y|X_j = x_j \sim N(\rho_j x, (1-\rho_j^2))$.  Let $$
G(y) := \Phi(x) \Phi(y) - \int_{-\infty}^y \Phi \Big( \frac{x- \rho_j u}{\sqrt{1-\rho_j^2}} \Big) \phi(u) du.
$$
Then we have 
$$
k_j(x) = 
%\frac{1}{\Phi(x)(1-\Phi(x))} \sup_y \left|\Phi(x) \Phi(y) - \int_{-\infty}^y \Phi \Big( \frac{x- \rho_j u}{\sqrt{1-\rho_j^2}} \Big) \phi(u) du\right| =: 
\frac{1}{\Phi(x)(1-\Phi(x))} \sup_y | G(y)|.
$$
Note that 
$
\frac{\partial G}{\partial y} = \Phi(x) \phi(y) - \Phi \Big(\frac{x-\rho_j y}{\sqrt{1-\rho_j^2}} \Big) \phi(y) = \phi(y) \left(\Phi(x)-\Phi \Big(\frac{x-\rho_j y}{\sqrt{1-\rho_j^2}} \Big) \right) 
$, 
which gives $
\frac{\partial G}{\partial y} \Big\vert_{y = y^*} = 0. 
$ where $y^* = x \big(\frac{1-\sqrt{1-\rho_j^2}}{\rho_j} \big)$.
When $\rho_j < 0$ then $G''(y^*) = \phi(y^*) \frac{\rho_j}{\sqrt{1-\rho_j^2}} \phi (x) < 0.$
Thus when $\rho_j<0$ then $G$ attains its supremum at $y=y^*$. Similarly, when $\rho_j>0$ then $-G$ attains its supremum at $y=y^*$.
It follows that  
\begin{align*}
k_j(x) &= \frac{1}{\Phi(x)(1-\Phi(x))} \left|\int_{-\infty}^{y^*} \Phi \Big( \frac{x- \rho_j u}{\sqrt{1-\rho_j^2}} \Big) \phi(u) du - \Phi(x) \Phi(y^*) \right|. 
%\ \ \text{when $\rho_j>0$}\\ 
%&= \frac{1}{\Phi(x)(1-\Phi(x))} \left(\Phi(x) \Phi(y^*)-\int_{-\infty}^{y^*} \Phi \Big( \frac{x- \rho_j u}{\sqrt{1-\rho_j^2}} \Big) \phi(u) du  \right) \ \ \text{when $\rho_j<0$} \\
%&=0 \ \ \text{when $\rho=0$}\\
%&=|\rho| \ \ \text{when $\rho = \pm 1$}
\end{align*}
When $\rho_j=1$, then 
$k_j(x) = \frac{1}{\Phi(x)(1-\Phi(x))} \left| \Phi(x) - \Phi(x) \Phi(x)\right| = 1.
$ When $\rho_j = -1$, then
$
k_j(x) = \frac{1}{\Phi(x)(1-\Phi(x))} \left| -\Phi(x)\Phi(-x)\right| = 1.
$

%Incidentally we know that $k_j(x) = 1$ when $\rho_j=1$ and $k_j(x) = 0$ when $\rho_j = 0$.
Now we show that $k_j(x)$ is an increasing function of $|\rho_j|$ by taking derivative $k_j(x)$ with respect to $\rho_j$. After some tedious calculations,
\begin{align*}
 \frac{\partial k_j(x)}{\partial \rho_j} &=\frac{\text{sgn}(\rho_j)}{\Phi(x)(1-\Phi(x))} (1-\rho_j^2)^{-3/2} \int_{-\infty}^{y^*} (\rho_j x - u) \phi\Big(\frac{x-\rho_j u}{\sqrt{1-\rho_j^2}} \Big) \phi(u)du\\
 &=\frac{\text{sgn}(\rho_j)}{\Phi(x)(1-\Phi(x))}\frac{(1-\rho_j^2)^{-1/2}}{2\pi} \exp \left(-x^2 h(\rho_j)\right),
\end{align*}
where $h(\rho_j) = \frac{1-\sqrt{1-\rho_j^2}}{\rho_j^2}$. Thus, $k_j(x)$ is increasing in $|\rho_j|$ since $h$ is symmetric.

\section{Proof of Theorem \ref{mainthm}}
Under the event that $\max_{j\in \{1,\ldots,p\}} | \hat K_j - K_j| < \Delta_S$, we know that 
\begin{align*}
\hat K_j > K_j - \Delta_S \geq \min_{j \in S} K_j - \Delta_S &= \max_{j \notin S} K_j, \ \ j \in S \\
\hat K_j < K_j + \Delta_S \leq \max_{j \notin S} K_j + \Delta_S &= \min_{j \in S} K_j, \ \ j \notin S.  
\end{align*}
Hence, for any $d_n \geq |S|$, we have $\hat S(|S|) \subset \hat S(d_n)$, which implies $S^* \subseteq \hat S(d_n)$.
On the other hand, by the following Lemma B.1, we have for any $\Delta_S > 4/n$,
\begin{align*}
P \left( \max_{j\in \{1,\ldots,p\}} | \hat K_j - K_j | \geq \Delta_S \right) &\geq 1- \sum_{j=1}^p P\left(|\hat K_j - K_j | \geq \Delta_S \right)\\ 
&\geq 1- p \left(4n\exp(-n\Delta_S^2/128) + 2\exp(-n\Delta_S^2/16)\right).
\end{align*}
It follows that when $\Delta_S \gg \sqrt{\log(pn)/n}$, the probability tends to 1.

\noindent
\textbf{Lemma B.1}
Consider $K_j$ in (\ref{cum.kf}) and $\hat K_j$ in (\ref{est.cum.kf}). Then for any $\epsilon>4/n$,
%Let $C > 1/4$. For $\epsilon \geq C\sqrt{\frac{\log n}{n}}$, and for $n$ large such that $n^{-1/4} (\log n)^{1/4} \leq 1/2C$,
$$P \left(| \hat K_j - K_j | \geq \epsilon \right) \leq 4n\exp(-n\epsilon^2/128) + 2\exp(-n\epsilon^2/16).
$$

\noindent
\textit{Proof of Lemma B.1}
Without loss of generality, we only need to consider $\epsilon < 1$ since otherwise, the probability in the left side is trivially 0. Also we assume that all $X_{\ell j}$ are distinct for convenience.
First, we use a simple triangle inequality to bound
\begin{align}
 P \left(| \hat K_j - K_j | \geq \epsilon \right) &= P \Big( \Big|\frac{1}{n}\sum_{\ell} \hat k_j(X_{\ell j}) - \frac{1}{n} \sum_{\ell} k_j(X_{\ell j}) +   \frac{1}{n} \sum_{\ell} k_j(X_{\ell j}) - E k_j (\tilde X_j) \Big| \geq \epsilon \Big) \nonumber \\
&  \leq P \Big(\Big|\sum_{\ell} \hat k_j(X_{\ell j}) - \sum_{\ell} k_j(X_{\ell j})\Big| \geq \frac{n\epsilon}{2} \Big) + \nonumber \\
& \qquad  \qquad  P \Big(\Big| \frac{1}{n} \sum_{\ell} k_j(X_{\ell j}) - E k_j (\tilde X_j) \Big| \geq \frac{\epsilon}{2} \Big) \nonumber \\
 &:= (i) + (ii). \label{twoterms}
\end{align}
{Then we treat the second term $(ii)$. By the Bernstein's inequality(e.g. Lemma 2.2.9 in \cite{vaart1996}), and using the fact that each $X_{\ell j}$ for $\ell=1,\ldots, n$ is independent and has the same distribution as the distribution of $\tilde X_{j}$, we have}
\begin{align*}
 (ii) &\leq 2\exp \left(-\frac{1}{8} \frac{n^2 \epsilon^2}{n+ n\epsilon/3} \right) \leq 2\exp \left(-\frac{1}{16} {n \epsilon^2}\right)
\end{align*}
where the first inequality follows by bounding the variance of each $k_j (X_{\ell j}) - E k_j(\tilde X_j)$ by 1 from the fact that $|k_j (X_{\ell j}) - E k_j(\tilde X_j)| \leq 1$ for any $\ell =1, \ldots n$. 

Now we consider the first term $(i)$ in (\ref{twoterms}). 
First note that $|\hat k_j(X_{\ell j}) - k_j(X_{\ell j})| \leq 1$ for any $\ell=1, \ldots, n$. We use this trivial bound for $\ell=\ell'$ where $X_{\ell' j}$ is the maximum of $X_{1j},\ldots,X_{nj}$. 
Let $\epsilon' := \epsilon/2-1/n$ and $\epsilon_\ell := \frac{1}{2} \sqrt{\frac{n}{\ell}}\epsilon'$. Using
$\sum_{\ell=1}^n \epsilon_\ell = \sum_{\ell =1}^n \frac{1}{2}\sqrt{\frac{n}{\ell}}\epsilon' \leq \frac{\sqrt{n}\epsilon'}{2} \int_{1}^n x^{-1/2}dx \leq n\epsilon',
$
we have by the union bound that 
\begin{align}
(i) = P \Big(\Big|\sum_{\ell=1}^n \hat k_j(X_{\ell j})- \sum_{\ell=1}^n k_j(X_{\ell j})\Big| \geq \frac{n\epsilon}{2} \Big) 
&\leq P \Big(\Big|\sum_{\ell \neq \ell'} \hat k_j(X_{\ell j})- \sum_{\ell \neq \ell'} k_j(X_{\ell j})\Big| \geq n\epsilon' \Big) \nonumber \\
%&\leq P \left(\Big|\frac{1}{n}\sum_{\ell = a_n+1}^{n-a_n} \Big( \hat k_j(X_{\ell j})-k_j(X_{\ell j})\Big) \Big| \geq \epsilon' \right) \nonumber \\
\leq \sum_{\ell \neq \ell'} {P} \Big( \Big|\hat k_j(X_{\ell j}) &- k_j(X_{\ell j}) \Big| \geq \epsilon_{\tilde \ell} \Big) \label{term}
\end{align}
where $\tilde \ell$ corresponds to the rank of $X_{\ell j}$.

We bound (\ref{term}) by above using similar ideas in Lemma A1 of \citet{mai2012kolmogorov}. Using the Dvoretzky--Kiefer--Wolfowitz inequality, for any $x$ in the support of $X_j$,
$$
{P} \left( \Big|\hat k_j(x) - k_j(x) \Big| \geq \epsilon_{\tilde \ell}  \Big\vert X_{1j}, \ldots, X_{nj} \right)\leq 2 \exp(-n_{+} \epsilon_{\tilde \ell}^2/2) + 2\exp(- n_{-} \epsilon_{\tilde \ell}^2/2)
$$
where $n_{+} = \sum_{i=1}^n \one_{\{X_{ij} >x\}}$ and $n_{-} = \sum_{i=1}^n \one_{\{X_{ij} \leq x\}}$.
Thus by replacing $x$ by $X_{\ell j}$ followed by taking the expectation, we have
$$
\sum_{\ell \neq \ell'} {P} \left( \Big|\hat k_j(X_{\ell j}) - k_j(X_{\ell j}) \Big| \geq \epsilon_{\tilde \ell}  \right)\leq \sum_{\ell=1}^{n-1} \Big( 2 \exp(-(n-\ell)\epsilon_\ell^2/2) + 2\exp(- \ell\epsilon_\ell^2/2) \Big).
$$

% Thus using $E e^{tB}  = (1-p+pe^t)^n$ where $B$ is a binomial random variables having $p$, and using $\epsilon_\ell \leq 1$ (by assuming $n$ large),
% \begin{align*}
% P \left( \Big|\hat k_j(X_{\ell j}) - k_j(X_{\ell j}) \Big| \geq \epsilon_\ell \right)  &\leq 2 E \left[ \exp(-n_{\ell,+} \epsilon_\ell^2/2) +\exp(- n_{\ell,-} \epsilon_\ell^2/2) \right] \\
% &=2 \left( 1- \ell/n + \ell/n \exp(-\epsilon_\ell^2/2) \right)^n + 2 \left( \ell/n + (1-\ell/n) \exp(-\epsilon_\ell^2/2)\right)^n \\
% &\leq 2 \exp\left(-\ell(1-e^{-\epsilon_\ell^2/2})\right)+ 2 \exp(-n\epsilon_\ell^2/2)\exp \left( \ell(e^{\epsilon_\ell^2/2}-1) \right) \\
% &\leq 2   \left[ \exp \left( - \ell \epsilon_\ell^2/4 \right) + \exp \left( - (n-\ell) \epsilon_\ell^2/4\right) \right]
% \end{align*}
% where the penultimate inequality follows by $e^x \geq \Big( 1+ \frac{x}{n} \Big)^n$, and the final inequality follows since $\exp(x) \leq 1+ x+ x^2$ where $|x| \leq 1$, and using $n> \ell (1+\epsilon_\ell^2/2)$. 
It follows by symmetry
\begin{align*}
(i) \leq 4 \sum_{\ell=1}^{n-1} \exp \left( - \ell \epsilon_\ell^2/2 \right) 
= 4 \sum_{\ell =1}^{n-1} \exp( -n\epsilon'^2/8)
\leq 4 n \exp(-n\epsilon^2/128),
\end{align*}
where the last inequality holds since $\epsilon' = \epsilon/2-1/n \geq \epsilon/4$.
The proof is complete.

\end{document}